\documentclass{aastex6}
\usepackage{epsfig}
\usepackage{url}
\usepackage{hyperref}
\usepackage{amsmath}
\usepackage{xspace}
\newcommand{\kms}{km~s$^{-1}$\xspace}

\begin{document}
	
	\shorttitle{Multi-Flux-Rope System} %
	
	\shortauthors{Awasthi et al.}
	
	\title{Pre-Eruptive Magnetic Reconnection within a Multi-Flux-Rope System in the Solar Corona} 
	
	\author{Arun Kumar Awasthi\altaffilmark{1}, Rui Liu\altaffilmark{1}, Haimin Wang\altaffilmark{2,3}, Yuming Wang\altaffilmark{1,4}, Chenglong Shen\altaffilmark{1,4}}
	
	\altaffiltext{1}{CAS Key Laboratory of Geospace Environment, Department of Geophysics and Planetary Sciences, University of Science and Technology of China, Hefei 230026, China; rliu@ustc.edu.cn}
	
	\altaffiltext{2}{Space Weather Research Laboratory, New Jersey Institute of Technology, University Heights, Newark, NJ 07102-1982, USA}
	
	\altaffiltext{3}{Big Bear Solar Observatory, New Jersey Institute of Technology, 40386 North Shore Lane, Big Bear City, CA 92314-9672, USA}
	
	\altaffiltext{4}{Synergetic Innovation Center of Quantum Information \& Quantum Physics, University of Science and Technology of China, Hefei 230026, China.}

\begin{abstract}
The solar corona is frequently disrupted by coronal mass ejections (CMEs), whose core structure is believed to be a flux rope made of helical magnetic field. This has become a ``standard'' picture although it remains elusive how the flux rope forms and evolves toward eruption. While 1/3 of the ejecta passing through spacecrafts demonstrate a flux-rope structure, the rest have complex magnetic fields. Are they originating from a coherent flux rope, too? Here we investigate the source region of a complex ejecta, focusing on a flare precursor with definitive signatures of magnetic reconnection, i.e., nonthermal electrons, flaring plasma, and bi-directional outflowing blobs. Aided by nonlinear force-free field modeling, we conclude that the reconnection occurs within a system of multiple braided flux ropes with different degree of coherency. The observation signifies the importance of internal structure and dynamics in understanding CMEs and in predicting their impacts on Earth.
\end{abstract}

\section*{Introduction}
With size on the order of solar radius, a coronal mass ejection \citep[CME; see the recent review by][]{Webb&Howard2012} releases $10^{30}$--$10^{33}$ ergs of magnetic energy within $\sim$10$^3$ seconds via magnetic reconnection, a fundamental and ubiquitous physical process that ``cuts and pastes'' field lines at localized field discontinuities, i.e., current sheets, in plasma \citep{Priest&Forbes2000}. During reconnection, magnetic free energy is rapidly converted into thermal and kinetic energies of bulk plasma while particles are accelerated to relativistic speeds. The energy release takes on three phases, namely, precursor, impulsive and gradual phase. The latter two, jointly known as the main phase, have been studied in great detail, whereas precursor processes are poorly known owing to subtle activity and emission during this phase \citep[e.g.,][]{Chifor2007, Awasthi2014, Wang2017}, yet they may provide critical information on the eruptive structure, which is extremely difficult to capture during the eruption when it evolves rapidly and the accompanying flare often emits intensely enough to saturate the CCD camera. 

Referring to a bundle of helical magnetic field lines, magnetic flux rope is considered the core structure of CMEs \citep[e.g.,][]{Forbes2000,Vourlidas2013} and is key to triggering the eruption if it loses equilibrium or suffers MHD instabilities \citep{Forbes2000,Forbes2006}. Flux ropes are also considered the building blocks of the solar atmosphere \citep{Rust2003} as magnetized plasma has a strong tendency to relax toward force-free helical equilibria through magnetic reconnection \citep{Taylor1986}. Indeed this fundamental structure exists ubiquitously in astrophysical and laboratory plasma, spanning a wide range of scales from ion inertial length in current sheets \citep{Loureiro&Udensky2016} to thousands of light years in astrophysical jets \citep{Marscher2008}. How flux ropes form in the solar atmosphere has been intensely debated. Leading theories depict the formation as a reconnection process between sheared field lines prior to \citep{Moore2001} or during the eruption \citep{Antiochos1999}, or, as a bodily emergence from below the photosphere \citep{Hood2009}. A coronal flux rope is often identified by its helical shape \citep[e.g.,][]{Rust&Kumar1996,Canfield1999,Liu2010,Zhang2012} because corona plasma is `frozen' into magnetic field, which is however extremely difficult to measure. The best one can do at present is to estimate the coronal field by extrapolating the vector fields at the surface. In a nonlinear force-free field (NLFFF) extrapolation, \citet{Liu2016} identified a flux rope by a coherent 3D region of enhanced twist number (with magnitude $\ge1$), the number of turns two neighboring field lines wind about each other. This region is enclosed by a thin quasi-separatrix layer \citep[QSL;][]{Demoulin2006}), separating the twisted rope from the surrounding, untwisted field. The rapid change in magnetic connectivity at QSLs is translated to high squashing factor \citep[typically $>100$;][]{Titov2002}. Such a coherent flux rope is prevalently adopted in models and numerical simulations, with helical field lines collectively winding about a common axis. 

However, interplanetary counterpart of CMEs (ICMEs) exhibit a wide range of magnetic structures, from an enhanced, smoothly-rotating magnetic field in magnetic clouds \citep{Burlaga1981}, to multiple magnetic clouds \citep[e.g.,][]{Wang2003}, to seemingly chaotic fields in complex ejecta \citep[e.g.,][]{Burlaga2002}. Magnetic clouds account for about 1/3 of ICMEs \citep{Chi2016}, the rest are too complicated to be modeled by a single flux rope. A complex ejecta may result from several interacting ICMEs \citep{Burlaga2002}, or directly from an inherently complex CME, as demonstrated occasionally in numerical experiments \citep{Lynch2008}. These largely remain speculations, because of our ignorance about the internal structure of coronal flux ropes, otherwise we may better predict whether an ICME would have strong and sustaining southward field, one of the most decisive factors inducing intense geomagnetic storms \citep[disturbance storm time index $\mathrm{Dst}\le-100$ nT;][]{Gonzalez2007,Shen2017}. A few recent studies start to touch on this important issue. \citet{Wang2017nc} inferred a non-uniform twist profile from the dynamic formation of a coronal flux rope, whose highly twisted core forms earlier than the less twisted outer shells. \citet{Liu2012} and \citet{Kliem2014} studied a ``double-decker" flux rope with two branches separated in altitude but sharing the same footpoint regions. Their studies focus on discrete transfer of flux and mass from the lower to higher branch resulting in the latter's eruption. In this regard, preflare activities may provide crucial insight into the key conditions and parameters leading to the eruptive processes. 

Here, the emission activities during a small flare before an imminent major eruption allow us to catch an important glimpse of the internal structure and dynamics of the pre-eruptive system, which has multiple flux-rope branches manifesting inter-branch braiding and magnetic reconnection.  In the sections that follow, we introduced the methods of data reduction in Section \ref{sec:data}, presented and analyzed the observations in Section \ref{sec:observation}, made concluding remarks in Section \ref{sec:conclusion}. 

\section{Instruments and Data Reduction} \label{sec:data}

\subsection{Processing of multi-wavelength images}
We analyzed the EUV images obtained by Atmospheric Imaging Assembly \citep[AIA;][]{Lemen2012} onboard \textit{Solar Dynamics Observatory (SDO)}. AIA provides un-interrupted observations of the full-disk Sun with a pixel size of $0''.6$ and a cadence of 12 s. The AIA's six EUV passbands in 94, 131, 171, 193, 211, and 335~{\AA} have distinctive temperature responses and cover a wide temperature range from 0.5--30 MK, which enables us to reconstruct the temperature distribution of plasma emitting along the line of sight in the optically thin corona, known as differential emission measure (DEM). Here we employed an algorithm providing positive definite DEM solutions by solving a linear system based on the concept of sparsity \citep{Cheung2015}.

The DEM-weighted mean temperature ($T_\mathrm{DEM}$) is defined conventionally as follows,
\begin{equation}\label{blob_t_em}
T_\mathrm{DEM}=\frac{\sum \mathrm{DEM}(T) \times T}{\sum \mathrm{DEM}(T)}, 
\end{equation}
which gives the total emission measure $\mathrm{EM}=\sum \mathrm{DEM}(T)\,dT$, with a binning $d\log T=0.1$ in this study. The thermal energy content in a region of interest is  
\begin{equation}\label{th_energy}
E_{th}=3k_BT\sqrt{\mathrm{EM}\,f V}
\end{equation}
where $k_B$ is the Boltzmann constant, $f$ denotes the filling factor assumed to be unity in this study, and $V$ the volume of plasma.

For the morphological investigation we mainly used three passbands, i.e., 131~{\AA} (\ion{Fe}{21} with peak response temperature $\log T = 7.05$; \ion{Fe}{8}, $\log T = 5.6$), 171~{\AA} (\ion{Fe}{9}, $\log T = 5.85$), and 304~{\AA} (\ion{He}{2}, $\log T = 4.7$). To highlight the braided structure in the EUV images, we applied the unsharp masking technique on 131~{\AA} images: a pseudo background image (the mask) is generated by smoothing the original image with a box-car of 10~$\times$~10 pixels ($6''\times6''$); the enhanced image is obtained by subtracting the background from the original image.

We also analyzed \ion{Ca}{2} images obtained by the Solar Optical Telescope (SOT) onboard \textit{Hinode} \citep{Kosugi2007} and \ion{Si}{4} 1400~{\AA} images from Interface Region Imaging Spectrograph \cite[IRIS;][]{DePontieu2014} to investigate the lower atmosphere response to the energy release during the flare. 

\subsection{Hard X-ray Imaging, Spectroscopy and Flare energetics}
The hard X-ray (HXR) emission from the flaring region is recorded by \textit{Rueven Ramaty High-Energy Solar Spectroscopic Imager} \citep[\textit{RHESSI};][]{Lin2002}). We synthesized HXR images with detectors 1, 3, 5, 6, 7, and 9, employing the PIXON algorithm \citep{Hurford2002}. Because RHESSI crosses the South Atlantic Anomaly (SAA), HXR data is only available from 16:29 UT on 2015 June 22. We prepared HXR spectra during 16:29 UT--16:52 UT with a time bin of 32 sec, and then performed forward fitting with a theoretical photon spectrum combining iso-thermal and thick-target bremsstrahlung models available in the SPectral EXecutive (SPEX) package within the SolarSoftWare (SSW) distribution\footnote{http://www.lmsal.com/solarsoft/}. The fitting procedure aims to minimize the reduced $\chi^2$ value to unity by iterations. The best-fit theoretical spectrum thus provides thermal and non-thermal characteristics of the source plasma. We further derived the thermal and non-thermal energy content during the flare. The thermal energy released is calculated with Eq.~\ref{th_energy}. The volume $V$ of the flaring plasma is approximated to be $A^{3/2}$, where $A$ denotes the area enclosing pixels with $\mathrm{EM}>3\times10^{26}$ cm$^{-5}$, a number chosen by trial and error to best represent the emitting region in the EM maps of 5-20 MK, as derived from AIA data. The thermal energy is overestimated because we assume the unknown filling factor ($f$) to be unity. The energy available in non-thermal electrons is derived by employing the function \texttt{calc\_nontherm\_electron\_energy\_flux.pro} in SPEX, using parameters obtained from the thick-target fitting, namely, electron flux, negative spectral index, low- and high-energy cut-off.

\subsection{Field Extrapolation, Squashing Factor and Twist Number}
We studied the magnetic field configuration by examining magnetograms from Helioseismic Magnetic Imager \citep[HMI;][]{Scherrer2012} onboard SDO. To extrapolate the coronal magnetic field, we employed Space-Weather HMI Active Region Patches (data product of \texttt{hmi.sharp\_cea} series) vector magnetograms at 12-minute cadence. The vector magnetograms are pre-processed to best suit the force-free condition before being fed into the ``weighted optimization'' NLFFF code as the photospheric boundary \citep{Wiegelmann2006}. Here we built the NLFFF over a uniform grid of $840\times 452\times 452$ pixels (pixel size 0.36 Mm) and investigated magnetic connectivities by tracing field lines pointwise on the bottom of a tenfold-refined grid with a fourth-order Runge-Kutta method, using footpoint positions of field lines to calculate the squashing factor $Q$ \citep{Titov2002}. Simultaneously we mapped twist number $T_w$ by integrating the local density of $T_w$,  $\nabla\times\mathbf{B}\cdot\mathbf{B}/4\pi B^2$, along each field lines \citep{Liu2016}.

\section{Results} \label{sec:observation}
\subsection{Overview}
The C1.1-class flare of interest occurs at 16:45 UT (peak time) on 22 June 2015 in the NOAA active region 12371, located close to the disk-center (N13W14). This is a compact flare without causing any CME, also known as a simple-loop flare, to be differentiated from the classical two-ribbon flares containing numerous flaring loops. Following the C-class flare, two more episodes of precursor emission at 17:24 and 17:42 UT \citep{Wang2017} precede an M6.5-class flare at 18:23 UT \citep{Jing2016,Jing2017}, a major eruption associated with a full-halo CME observed by the Large Angle and Spectrometric Coronagraph Experiment onboard the Solar and Heliospheric Observatory. All the above mentioned activities happen in the close vicinity of the polarity inversion line (PIL) that separates two major sunspots of opposite polarity in the center of the active region (Figure~\ref{fig:fr_nlfff}). Around this major PIL, we found no significant flux emergence or cancellation, and no significant photospheric shearing or converging motions within 16 hrs before the C-class flare, hence we focus on the corona in our investigations as elaborated below.

\subsection{Magnetic field configuration}
\begin{figure}[htbp]
\begin{tabular}{c}
  \epsfig{file=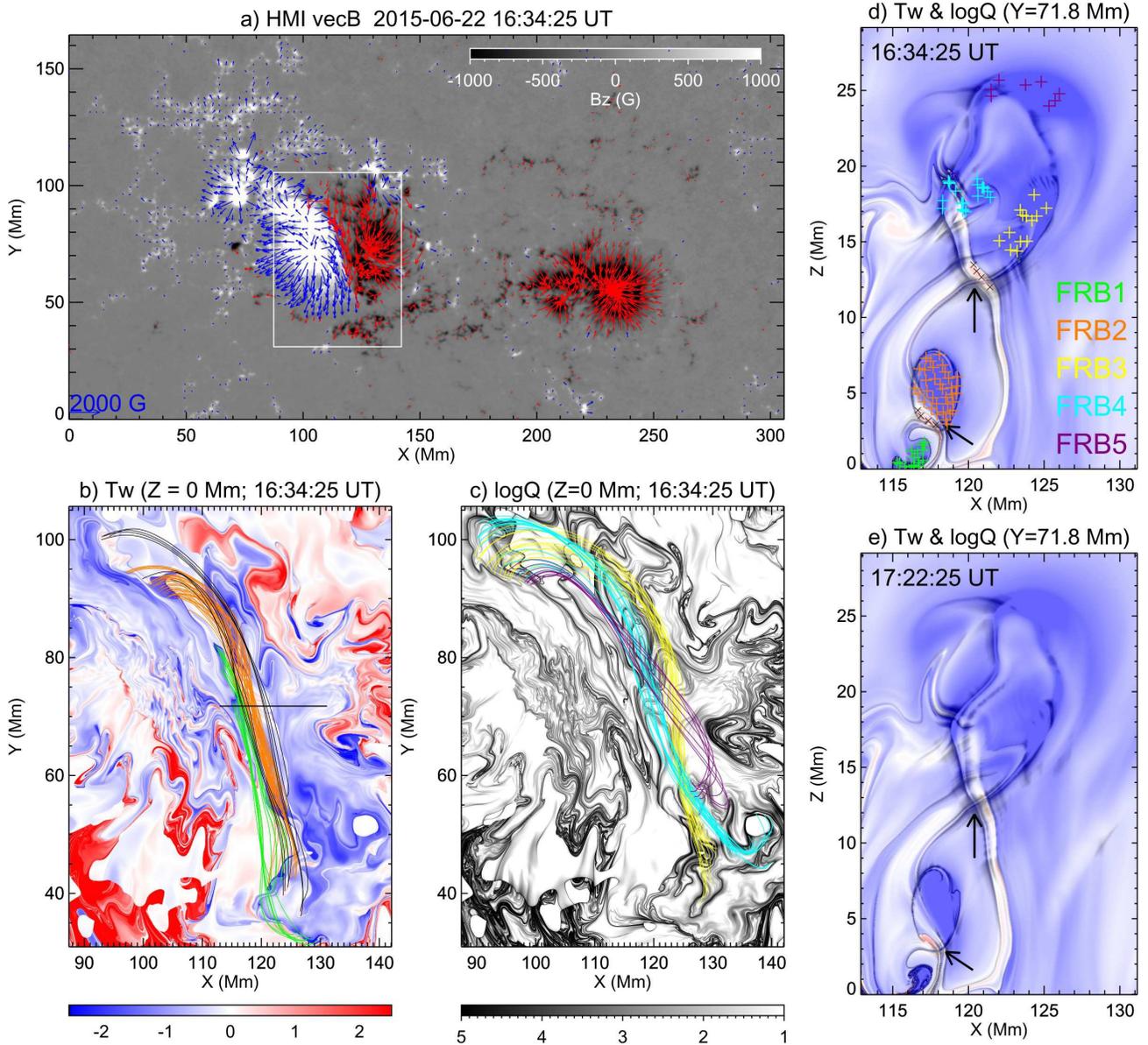, width=0.95\textwidth}
\end{tabular}
\caption{A complex flux-rope system as revealed by NLFFF modeling and field line tracing. [a]: Vector magnetogram of the active region obtained by HMI at 16:34:25 UT. $B_z$ component saturated at $\pm1000$G is shown as the background. The transverse field component is denoted by blue (red) arrows originating from positive (negative) polarity. The arrow at the bottom left corner gives the magnitude of arrows. [b]: Twist ($T_w$) map of a cut-out region shown by the white box in Panel [a]. The field lines in Panel [b] show two low-altitude flux-rope branches, FRB1 (green) and FRB2 (orange), while those in Panel [c] show three high-altitude branches, FRB3 (yellow), FRB4 (blue), and FRB5 (purple). [c]: $Q$-map ($\log Q$) of the same FOV as of Panel [b]. [d] \& [e]: Cross section of the flux rope in the $X$--$Z$ plane, denoted by a composite of $Q$ and $T_w$ map at 16:34:25 UT (precursor), and 17:22:25 UT (post flare), respectively. FRB3, FRB4, and FRB5 are bounded by a less coherent QSL than that delimiting FRB2. `+' (`x') symbols indicate where the twisted (sheared) field lines of the same color code in the Panels [b] \& [c] thread the plane. The two arrows in [d] and [e] point to two possible reconnection sites, where two QSLs intersects, displaying an X-type morphology. In the neighborhood, a region of positive twist (red) is noted. Subsequently, positive twist near the higher reconnection site disappears after the flare (Panel [e]; at 17:22:25 UT).}
\label{fig:fr_nlfff}
\end{figure}

\begin{figure}[htbp]
	\centering
	\begin{tabular}{c}
		\epsfig{file=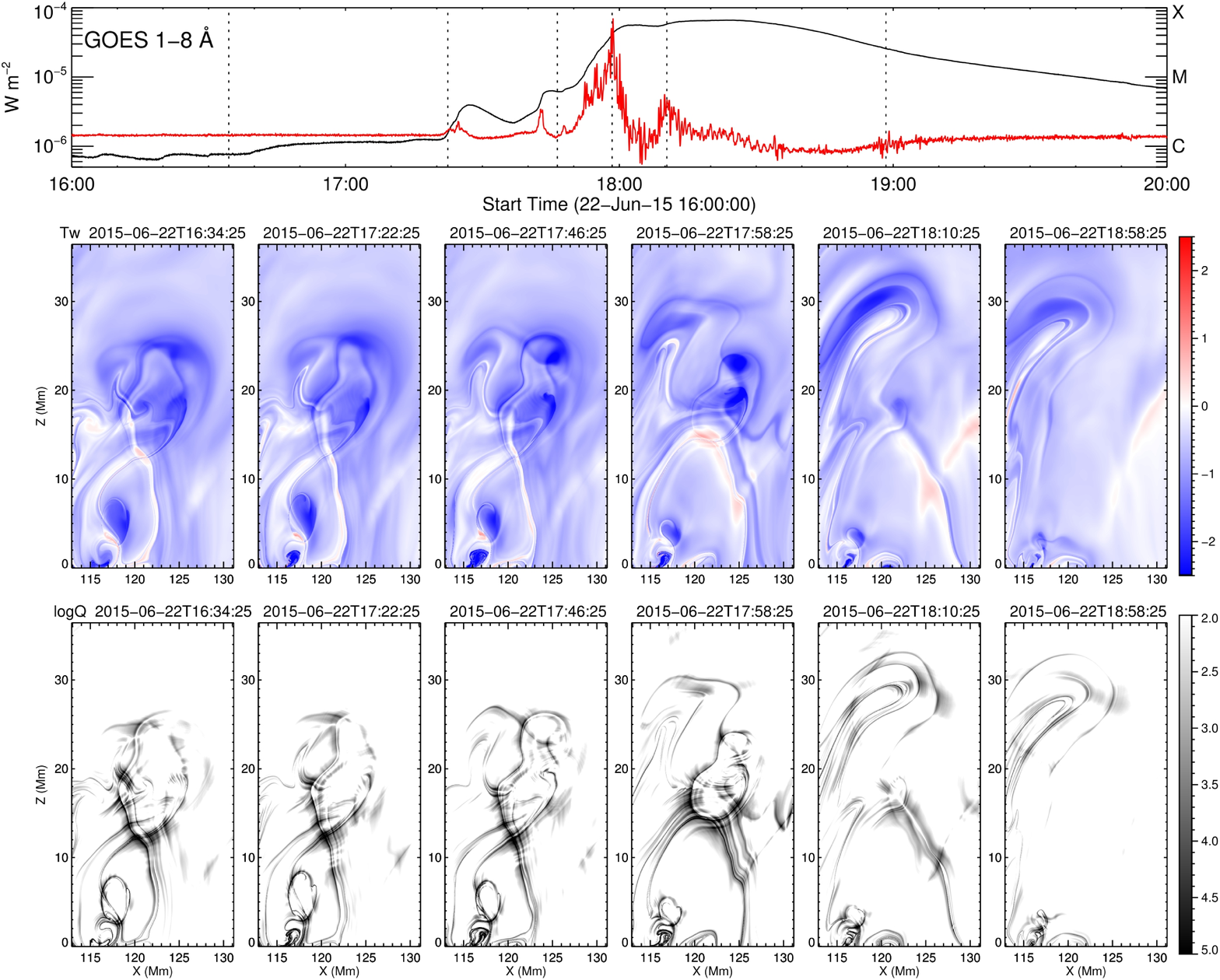, width=\textwidth}
	\end{tabular}
	\caption{Evolution of the multi-flux-rope system across the flares of interest. Top panel plots the GOES 1--8~{\AA} flux and its time derivative (red); dotted lines represent the time instances at which the maps of twist number $T_w$ (middle panels) and squashing factor $Q$ (bottom panels) are calculated.}
	\label{fig:mfr_evol}
\end{figure}

\begin{figure}[htbp]
	\centering
	\begin{tabular}{c}
		\epsfig{file=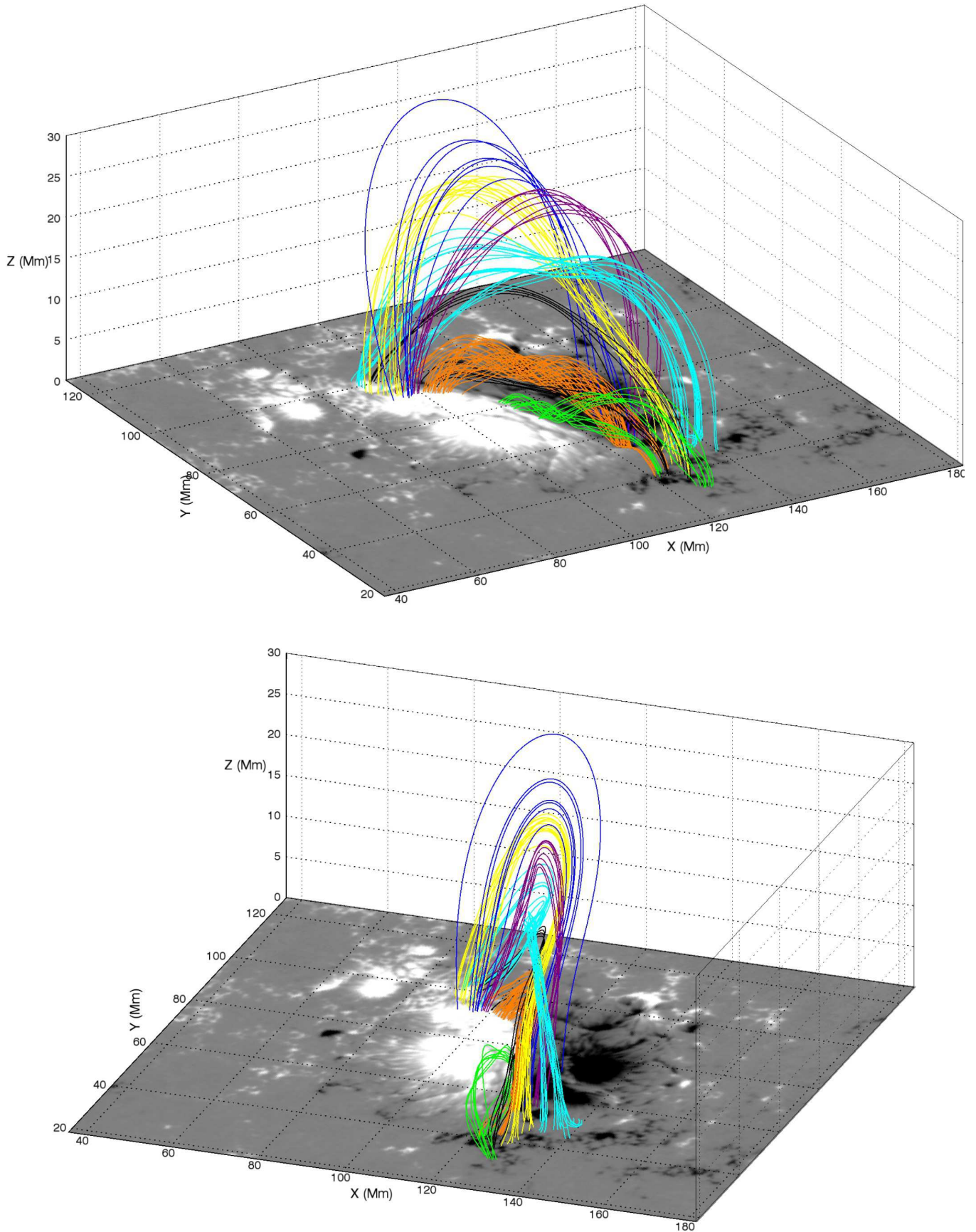, width=\textwidth}
	\end{tabular}
	\caption{Coronal field lines derived from the NLFFF at 16:34:25 UT in two different 3D perspectives. The field lines belonging to the five branches of the flux-rope system are shown in green, orange, cyan, yellow, and magenta. In blue are the overlying sheared loops, and in black the field lines of positive twist threading the intersection of QSLs (same color code as in Figure~\ref{fig:fr_nlfff}).}
	\label{fig:fl3d2}
\end{figure}

Magnetic-field restructuring during the precursor phase is understood to play a key role in triggering the impending flare \citep[e.g.,][]{Wang2017}. We employed NLFFF extrapolation method to model the coronal magnetic field, and selected a rectangular region covering the major PIL where the precursor emission is concentrated (Figure~\ref{fig:fr_nlfff}a) to derive the maps of squashing factor $Q$ and twist number $T_w$ (Figure~\ref{fig:fr_nlfff}(b and c)). A composite of $Q$ and $T_w$ maps in the X-Z plane at 16:34:25 UT and 17:22:25 UT, respectively, is plotted in Figure~\ref{fig:fr_nlfff}(d and e) (see also Figure~\ref{fig:mfr_evol}). Based on these maps and field-line tracing, we identified a system comprised of at least five flux-rope branches separated in altitude, labeled FRB1 (green), FRB2 (orange), FRB3 (yellow), FRB4 (cyan), and FRB5 (violet), following the sequence of low-to-high altitude. Each branch consists of twisted field lines displaying similar winding and footpoint regions. Suspended in the corona, FRB2 is a coherent rope displaying an oval with enhanced $T_w$ fully enclosed by a QSL, similar to the rope in \citet{Liu2016}, while the lower branch FRB1 is less coherent and apparently attached to the surface. Braiding with each other (Figure~\ref{fig:fl3d2}), the high-altitude set of the three branches (FRB3--5) are roughly bounded by a QSL (see also Figure~\ref{fig:mfr_evol}), which is not as well defined as the one enclosing FRB2. At 16:34:25 UT the combined map of $T_w$ and $Q$ (Figure~\ref{fig:fr_nlfff}d; see also Figure~\ref{fig:mfr_evol}) shows an opposite (positive) twist region beneath FRB3 and FRB2, where the two high-Q layers intersect, a favorable site for 3D magnetic reconnection \citep{Demoulin2006}. We traced a few representative field lines (black) threading these positive-twist regions, which have twist numbers of $0.41\pm0.14$. Compared with the post-flare map at 17:22:25 UT (Figure~\ref{fig:fr_nlfff}e and Figure~\ref{fig:mfr_evol}), the positive-twist region beneath FRB3 largely disappears, supposedly via the cancellation with the dominant negative twist. 

It is well known that reconnection-related changes in the coronal field can be noticed from comparing the NLFFF before and after \citep[e.g.,][]{Liu2016}, as NLFFF extrapolation reconstructs magnetic topology in active regions with high fidelity \citep[e.g.,][]{Liu2014,Liu2016SR} while magnetic reconnection changes topology. As following, we analyze multi-wavelength coronal observations to seek reconnection signatures, of which the most sought-after are plasma heating, reconnection outflows, and nonthermal particle acceleration.

\subsection{Signatures of Magnetic Reconnection} 
\begin{figure}[htbp]
\centering
\begin{tabular}{c}
    \epsfig{file=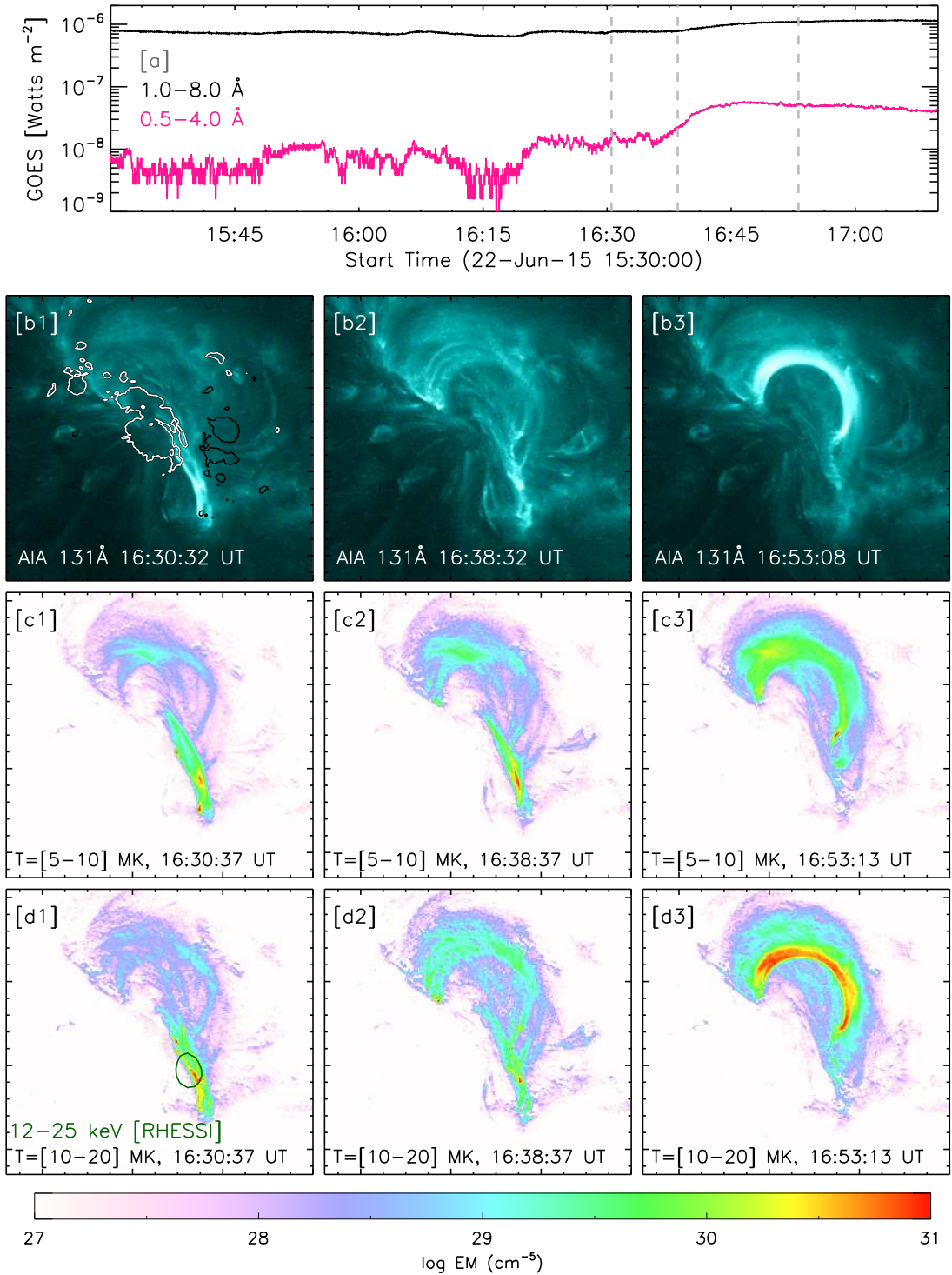, width=0.8\textwidth}
\end{tabular}
\caption{Morphological evolution and thermal characterization of the C1.1 flare under investigation. [a]: Temporal evolution of X-ray flux in 1-8 $\mathrm{\AA}$, and 0.5-4 $\mathrm{\AA}$ as recorded by the \textit{GOES} satellite. Dotted lines represent the times instances at which the snapshots presented below are taken. [b1-b3]: Time sequence of 131 $\mathrm{\AA}$ images recorded by AIA during the precursor ([b1]) and main phases ([b2] and [b3]) of the flare. Superimposed are contours of signed (positive in white and negative in black) magnetic-field strength at 800 G. Corresponding to the times instances of EUV images, EM maps are shown in the temperature range 5-10 MK (Panels [c1-c3]) and 10-20 MK (Panels [d1-d3]), respectively. A contour overplotted on [d1] represents 70\% of the maximum intensity of \textit{RHESSI} 12-25 keV image. An animation is available on-line.}
\label{fig:timeline_t_dem}
\end{figure}

Small, intermittent enhancements can be seen in the soft X-ray (SXR) 1--8 {\AA} lightcurve as early as $\sim\,$1 hour before the C-class flare at 16:34 UT (Figure~\ref{fig:timeline_t_dem}a), but we focused on activities from $\sim\,$16:18 UT when the emission level becomes persistently elevated. Snapshots of the flare during the precursor and main phase are shown in 131~{\AA} (Figure~\ref{fig:timeline_t_dem}; see also the accompanying movie). Superimposing line-of-sight component of magnetic field (contours) onto the 131~{\AA} image (Figure~\ref{fig:timeline_t_dem}(b1)), one can see that the EUV emission during the precursor phase is dominated by multiple threads apparently entangled and aligned along the PIL of interest (Figure~\ref{fig:timeline_t_dem}(b1)). The main phase of the flare is pronounced in the form of multiple overlying loops (Figure~\ref{fig:timeline_t_dem}(b2)), which evolve into a thick loop with enhanced emission during the gradual phase (Figure~\ref{fig:timeline_t_dem}(b3)). Further, we obtained the EM of the flaring plasma at 5-10 (Figure~\ref{fig:timeline_t_dem}(c1-c3)) and 10-20 MK (Figure~\ref{fig:timeline_t_dem}(d1-d3)), respectively. It is clear that the threads along the PIL are heated up to $\sim\,$20 MK during the precursor phase.

\begin{figure}[htbp]
\begin{tabular}{c}
  \epsfig{file=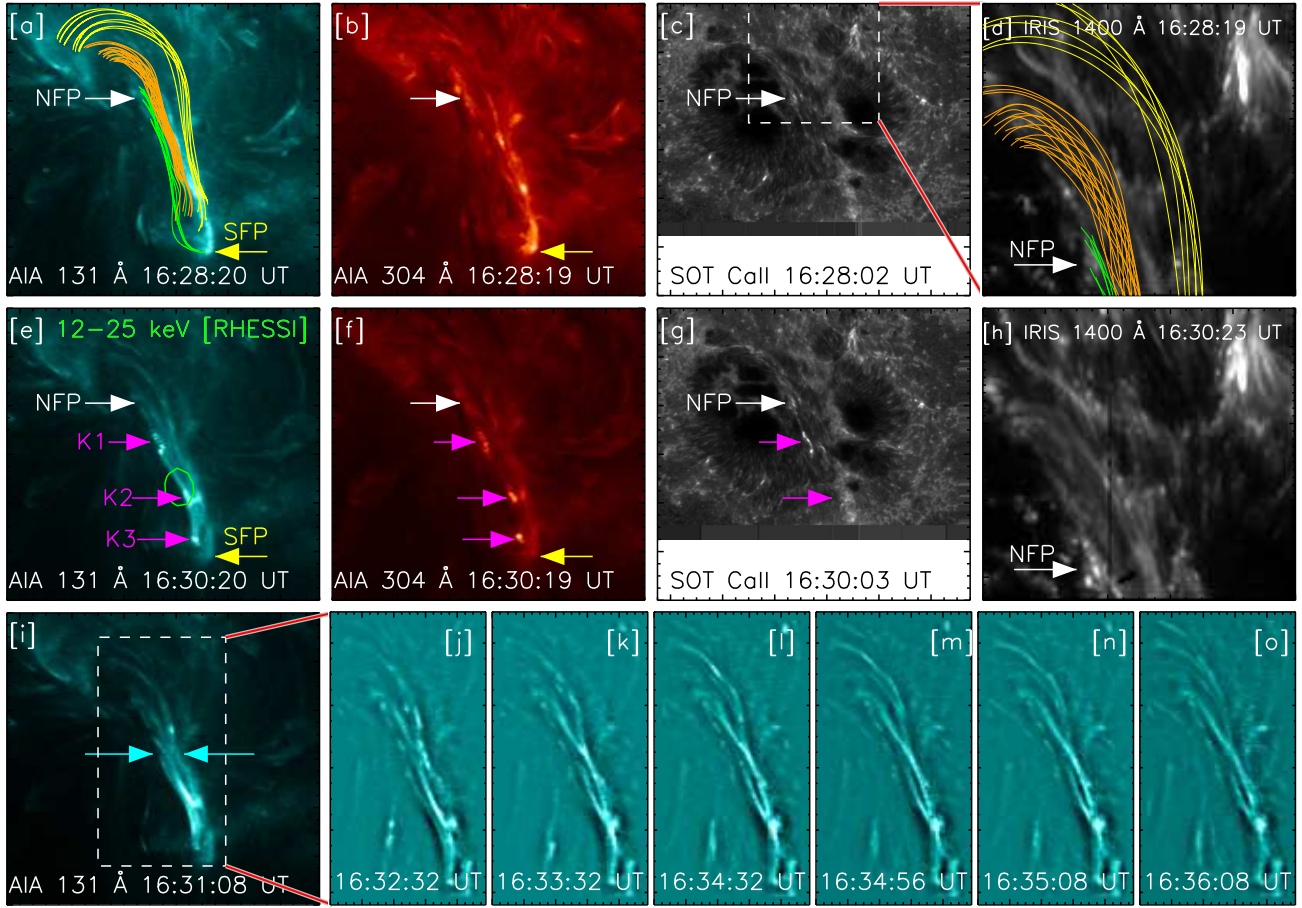, height=0.95\textwidth, angle=90}
\end{tabular}
\caption{Signature of distinctive energy release at various layers of the solar atmosphere. Panels [a]-[h]:  Multi-wavelength images obtained by AIA 131~{\AA} and 304~{\AA} passbands, SOT \ion{Ca}{2}, and IRIS \ion{Si}{4} 1400~{\AA}, shown from left to right. Field lines corresponding to three flux-rope branches, namely FRB1, FRB2 and FRB3 are drawn in Panels [a] and [d]. The dashed box in Ca II image (Panel [c]) corresponds to the FOV of the IRIS images ([d] \& [h]). NFP (SFP) denoted the brightening corresponding to the northern (southern) footpoints of the flux-rope system, K1-K3 represents various brightened kernels co-spatial to the spine of the flux-rope system. A \textit{RHESSI} 12-25 keV emission contour is overplotted on Panel [e] denoting the reconnection site. Panels [i]--[o] show the braided threads in 131~{\AA} images,  crossing each other as marked by arrows in Panel [i]. A small section of the image (dotted box in Panel [i]) is enhanced by unsharp mask (Panels [j]--[o]) to highlight the fine structure of the braided threads.}
\label{fig:reconnection_obs}
\end{figure}

Overplotting the field lines of FRB1 (green), FRB2 (orange), and FRB3 (yellow; representing the high-altitude flux-rope branches for simplicity) on the 131~{\AA} image (Figure~\ref{fig:reconnection_obs}a), one can see the clear spatial association between the flux-rope branches and the entangled threads. The two ends of this system (labeled NFP and SFP) are associated with extended surface brightenings in AIA 304~{\AA}, IRIS 1400~{\AA} and SOT Ca II. At $\sim\,$16:30 UT (Figure~\ref{fig:reconnection_obs}(e-g)), three brightened emission kernels (labeled K1, K2 and K3) are seen at both 131 and 304~{\AA}. K1 is also seen in the Ca II image, suggesting that it is a footpoint emission in the low atmosphere. On the contrary, K2 is missing in the Ca II image, suggesting it occurs relatively high in the corona. These emission kernels are associated with the enhanced emission co-spatial to NFP, as distinctly seen in the 1400~{\AA} image (white arrow), as well as to SFP (yellow arrow), which is in agreement to the scenario of reconnection within different flux-rope branches: energy is released at the reconnection site as indicated by nonthermal hard X-ray (HXR) emission (see Figure~\ref{fig:reconnection_obs}e and below) and further deposited at the flux-rope footpoints. Further, in the wake of the reconnection episode, the bright threads become more braided than before, with some threads apparently crossing each other (Figure~\ref{fig:reconnection_obs}(i--o)), indicating an ongoing magnetic reconfiguration. The braiding may further contribute to energizing coronal plasma by the dissipation of currents induced by the entangled field lines \citep{Parker1983a}.

\begin{figure}[htbp]
\centering
\begin{tabular}{c}
  \epsfig{file=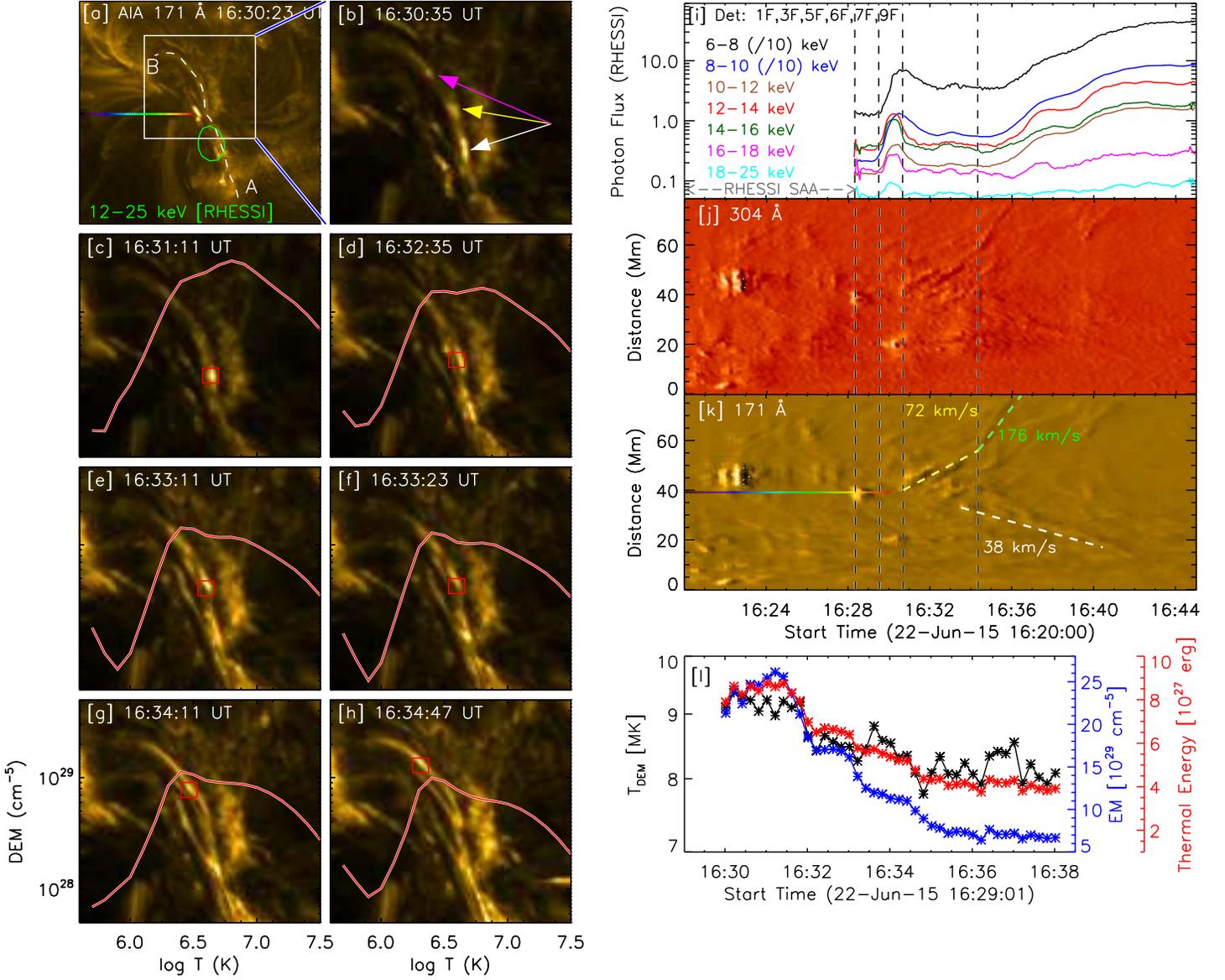, width=0.99\textwidth}
\end{tabular}
\vspace{2em}
\caption{Kinematics and thermodynamics of the outflowing blobs. [a]--[h] Sequence of AIA 171~{\AA} images. A contour of \textit{RHESSI} 12-25 keV image is overplotted on [a] to mark the reconnection site. Three representative blobs are denoted by arrows in [b]. One distinct blob, marked by a red box in [c]-[h], has been tracked as it propagates along the spinal direction of the flux-rope system. The blob's EM distribution is overplotted. Panel [i] shows \textit{RHESSI} lightcurves in various energy bands. Panels [j] and [k] shows the time-distance diagrams derived from the slit (white dotted-curve) in Panel [a] in the direction `A' to `B', using running difference images in 304 and 171~{\AA}, respectively. A rainbow colored line in Panel [a] corresponds to the reference line as marked in the time-distance diagrams. The speed of the blobs moving northward is estimated to be 72~\kms (increased to 176~\kms at the onset of the main phase of the flare) while that in the opposite direction is estimated to be 38~\kms. Panel [l] shows the DEM-weighted temperature (black; scaled by the left axis), EM (blue; scaled by the right axis), and thermal energy content (red; scaled by the rightmost axis) of the tracked blob (red square in [c]-[h]).}
\label{fig:recnn_outflow_evol}
\end{figure}

We further analyze the kinematics and thermal characteristics of several brightening features produced as a consequence of reconnection. Figure~\ref{fig:recnn_outflow_evol} presents snapshots of 171~{\AA} images representing the brightening activities followed by the reconnection episode as shown in Figure~\ref{fig:reconnection_obs}. The images reveal that some blobs originate from the close vicinity of the reconnection site as indicated by nonthermal HXR emission (see Figure~\ref{fig:recnn_outflow_evol}b and below) and propagate away along the spinal direction of the flux-rope system. Among many episodes of flows representing distinctive blobs, we focus on one prominent episode with the flow onsetting at the peak of a small, yet impulsive HXR bump at 16:30 UT. The flow speed of $\sim\,$72~\kms is estimated from the time-distance diagrams (Figure~\ref{fig:recnn_outflow_evol}(j and k)). These are made by taking slices off the running-difference images along the curved flow path (dashed curve in Figure~\ref{fig:recnn_outflow_evol}a) and then stacking them up chronologically. A further increase in speed to $\sim\,$176~\kms is noted prior to the onset of the flare main phase at 16:35 UT. The time-distance diagrams also reveal a counterflow at 38~\kms, whose lower speed is likely due to projection effects. We interpret this set of bi-directional flow as a signature of reconnection outflow.

To investigate the thermal characteristics of the reconnection outflows, we tracked one blob distinctly observed during 16:31--16:34 UT (marked by red boxes in Figure~\ref{fig:recnn_outflow_evol}) and solved DEM solutions over 0.5--30 MK within a 4$''$ wide box enclosing the blob. The DEM distribution reveals a hot component peaking at $\sim\,$10 MK due to reconnection-induced heating, and a `cold' component at $\sim\,$1.5 MK mainly attributed to the `quiet' corona in the foreground and background of the blob along the line of sight. From the evolution of DEM-weighted mean temperature $T_\mathrm{DEM}$ (Figure~\ref{fig:recnn_outflow_evol}) one can see that the blob temperature attains a maximum value $\sim\,$10 MK in the beginning and decays to $\sim\,$8 MK, as far as it can be identified, due presumably to cooling. Its emission measure (EM) peaks approximately one minute later than the temperature maximum and varies between [0.5--2.5]$\,\times10^{30}$ cm$^{-5}$. The thermal energy content of each individual blob is estimated by assuming its volume as a cube of width $4''$, which effectively encloses the blob. The results vary in the range of 4--9 $\times$ 10$^{27}$ ergs (Figure~\ref{fig:recnn_outflow_evol}[l]), amounts to a subflare or the largest nanoflare \citep{Parker1988}.

\begin{figure}[htbp]
\begin{tabular}{c}
  \epsfig{file=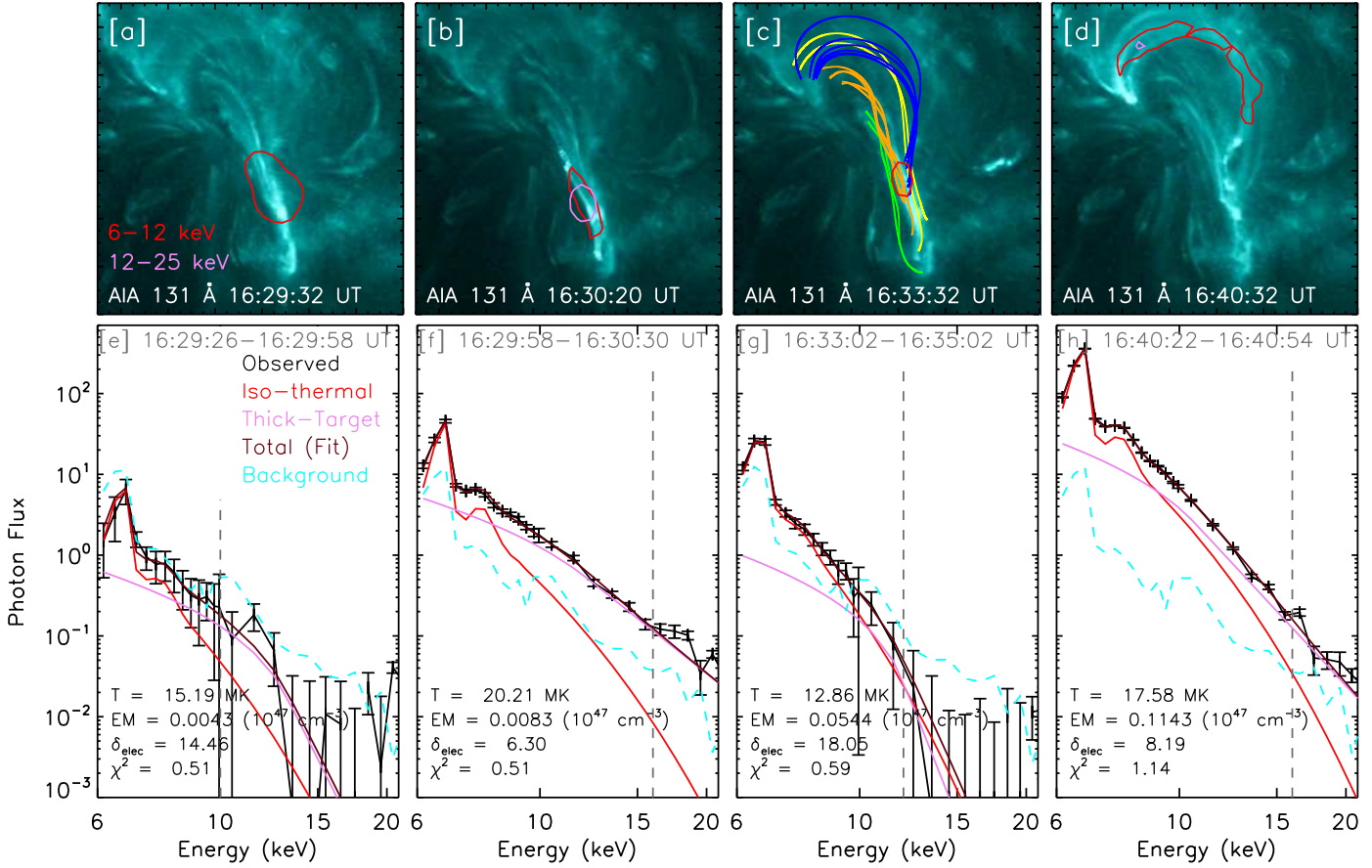, width=0.95\textwidth}
\end{tabular}
\caption{HXR diagnostics of the flaring plasma. [a]--[d]: Sequence of 131~{\AA} images, superimposed on which are the magnetic field lines of the flux-rope branches, FRB1 (green), FRB2 (orange), and FRB3 (yellow), and of the overlying loops (blue), as well as contours of the X-ray images corresponding to 60\% of the maximum emission in 6-12 keV (red) and 12-25 keV (magenta) energy bands. [e]--[h]: Background-subtracted photon spectrum (black) and the best-fit model (cyan) combining iso-thermal (red) and non-thermal thick-target bremsstrahlung (magenta). The vertical dotted line (grey) marks the high-energy limit of the fitting. The fitting parameters manifest the presence of high-temperature plasma and non-thermal electrons ($\delta=-6.3$), supporting the scenario of magnetic reconnection in the early stage of the flare evolution. }
\label{fig:euv_xray_evol}
\end{figure}

\begin{figure}[htbp]
\centering
\begin{tabular}{c}
    \epsfig{file=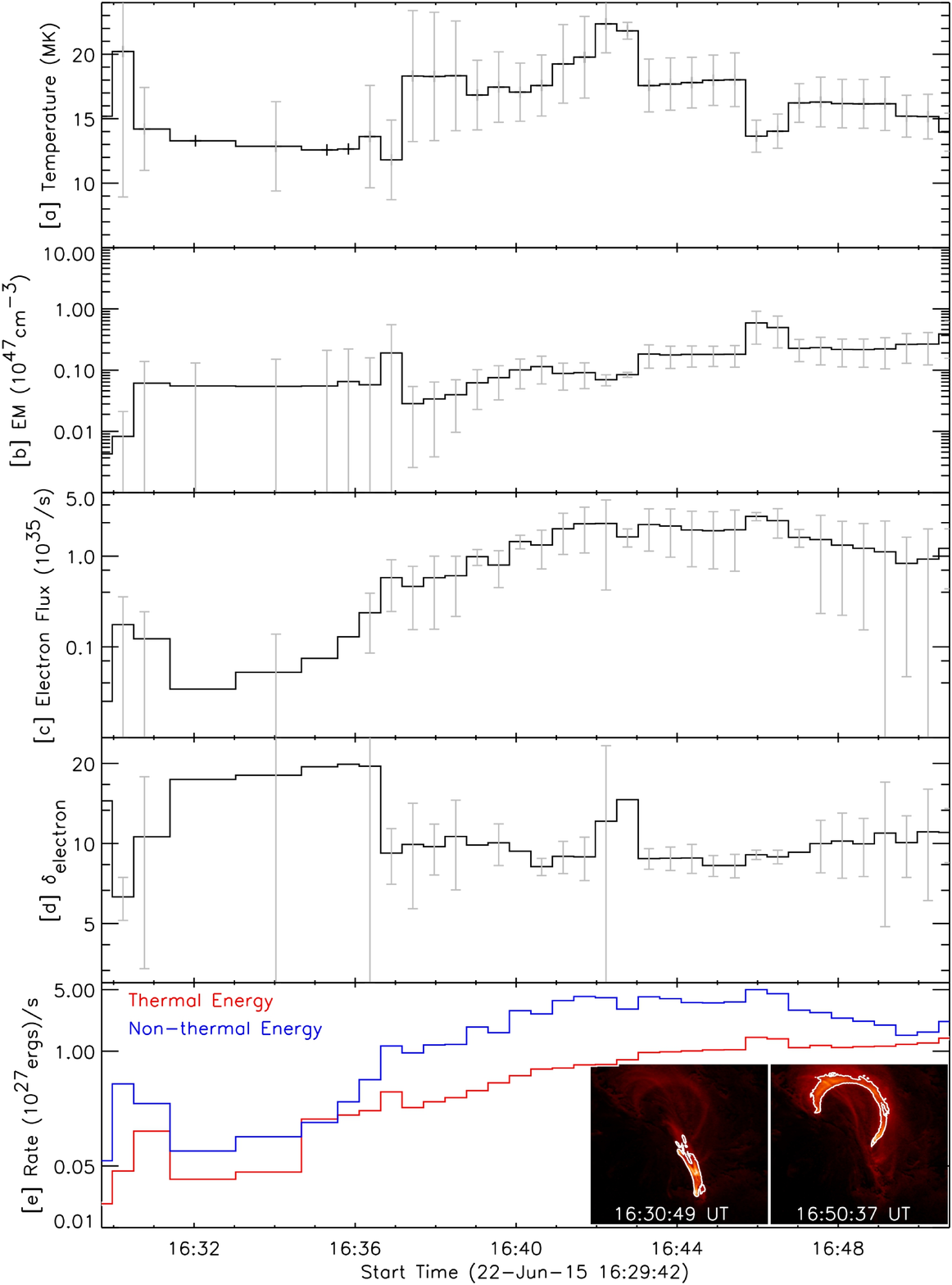, width=\textwidth}
\end{tabular}
\caption{Energetics of the flare plasma. [a]--[d] Temporal evolution of thermal and non-thermal characteristics of the flare plasma as derived from analyzing \textit{RHESSI} HXR spectra. [e] Thermal and non-thermal energy release rate during various phases of the flare. The insets show the EM maps of 5-20 MK for two instances. To estimate the thermal energy, the emitting plasma's volume is derived from the area within the contour of $\mathrm{EM}>3\times10^{26}$ cm$^{-5}$ drawn on the EM maps.}
\label{fig:hessi_thermal_nth_energy}
\end{figure}

The spatial and spectral evolution of HXR emission is studied in conjunction with the reconnection episode. Superimposing HXR sources over AIA 131~{\AA} images (Figure~\ref{fig:euv_xray_evol}(a--d)), one can see that during the precursor phase the HXR emission is co-spatial to the EUV enhancement in the center of the braided threads aligned along the PIL. The HXR source in 12--25 keV at 16:30 UT (Figure~\ref{fig:euv_xray_evol}b) corresponds to the power-law component of the photon spectrum (Figure~\ref{fig:euv_xray_evol}f). The spectral fitting reveals the presence of non-thermal electron flux with a spectral index ($\delta$) of 6.3 and of hot plasma at 20 MK, consistent with the DEM analysis. This precursor HXR emission also coincides in time with the onset of outflowing plasma blobs (Figure~\ref{fig:recnn_outflow_evol}). The bidirectional outflows and the presence of nonthermal electrons along with the high-temperature plasma argue strongly for the occurrence of magnetic reconnection within the flux-rope system, as the outflows are directed along its spinal direction. In contrast, during the main phase of the flare, the HXR spectra can be better fitted by an exponential function depicting thermal bremsstrahlung at a lower temperature than during the precursor phase (Figure~\ref{fig:euv_xray_evol}h), and the corresponding HXR source takes the shape of a thick loop similar to its EUV counterpart arching over the braided threads (Figure~\ref{fig:euv_xray_evol}d). We found that the nonthermal electron energy content is generally sufficient to energize the thermal emission during the flare (Figure~\ref{fig:hessi_thermal_nth_energy}), despite that the electron spectra is significantly harder during the precursor phase than other phases. The nonthermal energy released by reconnection within the flux-rope system is expected to be deposited at its footpoints, where the dense chromospheric plasma is heated and expands along the flux-rope field lines undergoing reconnection into the corona. We conjecture that these twisted field lines later relax into the less twisted, i.e., sheared field lines (blue; Figure~\ref{fig:euv_xray_evol}c) to produce the post-flare loops emitting thermal X-rays and EUV (Figure~\ref{fig:euv_xray_evol}d).

\subsection{Interplanetary Effects}
\begin{figure}[htbp]
\centering
\begin{tabular}{c}
    \epsfig{file=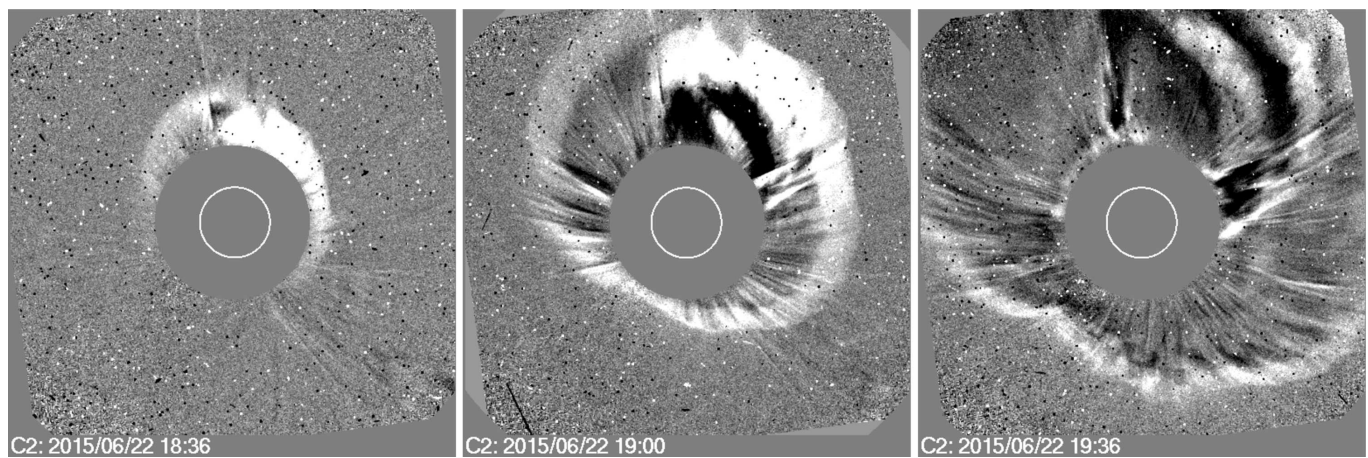, width=\textwidth}
\end{tabular}
\caption{Full-halo CME recorded by LASCO's C2 coronagraph (2.2--7 solar radii). The CME is associated with the M6.5 flare immediately after the C1.1 flare under investigation.}
\label{fig:cme_evol}
\end{figure}

\begin{figure}[htbp]
	\centering
	\begin{tabular}{c}
		\epsfig{file=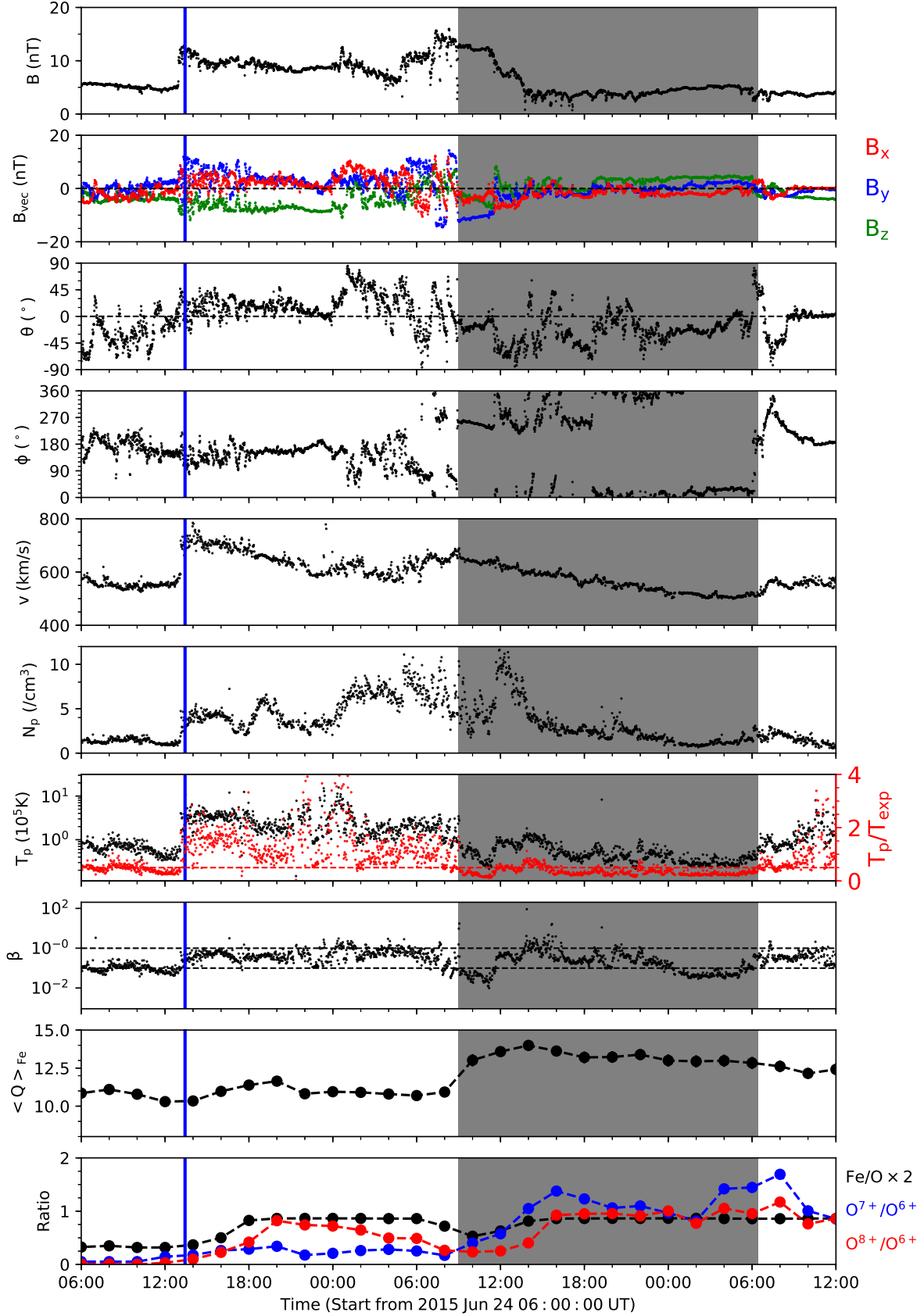, width=0.9\textwidth}
	\end{tabular}
	\vskip -10pt
	\caption{ICME observed by two near-Earth spacecrafts, Advanced Composition Explorer (ACE) and Wind. The shaded region during 2015 June 25--26 is identified as the interplanetary counterpart of the CME associated with the M6.5 flare that occurred on 22 June 2015, which is preceded by a shock (blue vertical line). From top to bottom are the magnetic field magnitude $B$, three components of the field in the geocentric solar ecliptic (GSE) coordinate system, field inclination angle $\theta$ (with respect to the ecliptic plane), azimuthal angle $\phi$ (0 deg pointing to the Sun), solar wind speed $V$, proton density $N_p$, proton temperature $T_p$ (superimposed by $T_p/T_\mathrm{exp}$), plasma $\beta$, average charge states of iron $\langle Q \rangle_\mathrm{Fe} = \sum_{i}Q_in_i$ (density is normalized such that $\sum_in_i = 1$), and various composition ratios. Data on magnetic field, ionic charge states, and composition are given by ACE, while data on bulk plasma by Wind, as the corresponding ACE data have large gaps.}
	\label{fig:ace}
\end{figure}

After the C-class flare, this multi-flux-rope system continues to evolve (Figure~\ref{fig:mfr_evol}) and erupts about 1 hr later as a full-halo CME propagating at about 1200 \kms in the outer corona (Figure~\ref{fig:cme_evol}). Such CMEs are known to be responsible for the majority of the most intense geomagnetic storms \citep{Webb&Howard2012}, but the current one only causes a moderate geomagnetic storm ($\mathrm{Dst_{min}}\approx-80$ nT), when it arrives at the Earth three days later as a shock-driven ejecta (Figure~\ref{fig:ace}). Its characteristics are typical of ICMEs \citep{Zurbuchen&Richardson2006}: the speed declines smoothly like a single stream expanding as a whole, the plasma $\beta$ (ratio of thermal and magnetic pressure) and proton temperature $T_p$ are depressed, so is $T_p/T_\mathrm{exp}$, indicating that the ejecta expands faster than the ambient solar wind, as $T_\mathrm{exp}$ is given by the well-established correlation between the solar-wind speed and temperature; on the other hand, the average Fe charge state $\langle Q\rangle_\mathrm{Fe}$ is enhanced, O$^{7+}$/O$^{6+}$ also shows a bump inside the ejecta. Frozen in as the CME expands into the outer corona, ``hot'' ionic charge states are reliable indicators of ICME plasma. However, the ejecta's magnetic field is very irregular, making it impossible to identify the individual components of the source structure. But it is unlikely that the ejecta could result from successive CMEs merging together, because it lasts only $\sim\,$22 hrs, the typical size of a single CME expected at 1 AU, which is distinct from those long-duration (typically over two days) events in which the CME-CME interaction is supposedly at play \citep{Burlaga2002,Wang2003}. In the LASCO CME catalog\footnote{\url{https://cdaw.gsfc.nasa.gov/CME_list/}} we found no candidate that had a fair likelihood to interact with the CME of interest. It is also highly unlikely that the spacecrafts only made a glancing encounter with this fast, earth-directed CME. Thus, the identity of the individual flux-rope branches at the Sun must have gradually lost as they continue to interact with each other and with the solar wind during the propagation from Sun to Earth. The resultant field irregularity explains why this CME causes no severe geo-effects. 

\section{Conclusion and Discussion} \label{sec:conclusion}
With multi-wavelength diagnostics of the flare precursor, our investigation has revealed for the very first time kinematic and thermodynamic characteristics of the reconnection within a multi-flux-rope system. We have identified definitive signatures of magnetic reconnection including significant flux of nonthermal electrons up to 20 keV, hot plasma up to $\sim$20 MK, and bi-directional outflows in the form of plasma blobs, which originate from the close vicinity of HXR emission and are directed along the spinal direction of the flux-rope system. These blobs have similar thermal and morphological characteristics as those associated with current sheets \citep[e.g.,][]{Takasao2012,Liu2013}. We conclude that the multi-flux-rope configuration together with internal interactions result in the complex ejecta observed in interplanetary space. Complex ejecta cannot be predicted by current models of CMEs, most of which include a single flux rope. The observations have shown that we cannot ignore the complexity of the pre-eruptive structure and the associated internal dynamics if we are to understand CMEs and their geo-effects. 

Compared with a single or double flux rope, what is novel and important is the braiding among the flux-rope branches, which introduces new degrees of freedom as well as additional free energy. Like entangled flux tubes \citep{Parker1983a}, entangled flux-rope branches are subject to internal reconnections at current sheets that form wherever two flux ropes are brought close enough together \citep[e.g.,][]{Linton2001}. While reconnections above or beneath a flux rope often lead to significant disturbance or even disruption \citep{Moore2006}, internal reconnections seem to favor plasma relaxation. This provides a mechanism for compact flares, which are the most numerous but not well understood within the frame of the standard model. With intermittent internal reconnections, however, the flux transferred from one rope to another may accumulate to the tipping point of eruption \citep[e.g.,][]{Su2011,Liu2012,Kliem2014,Zhang2014}. This could be the case in numerous observations that a major eruption is preceded by a few compact flares. Following the eruption, internal reconnections may continue to contribute to the strong heating in CMEs detected in the outer corona \citep{Akmal2001}.

\acknowledgments A.K.A. and R.L. are supported by NSFC 41474151, 41774150, and 41761134088. A.K.A. acknowledges the International postdoctoral program of USTC. H.W. is supported by US National Science Foundation. Y.W. acknowledges the support from NSFC 41774178 and 41574165. This work is also supported by NSFC 41421063, CAS Key Research Program of Frontier Sciences QYZDB-SSW-DQC015 and the fundamental research funds for the central universities.

%\bibliographystyle{apj}%{biblatex-nature}
%\bibliography{references}

\begin{thebibliography}{}
	\expandafter\ifx\csname natexlab\endcsname\relax\def\natexlab#1{#1}\fi
	
	\bibitem[{{Akmal} {et~al.}(2001){Akmal}, {Raymond}, {Vourlidas}, {Thompson},
		{Ciaravella}, {Ko}, {Uzzo}, \& {Wu}}]{Akmal2001}
	{Akmal}, A., {Raymond}, J.~C., {Vourlidas}, A., {et~al.} 2001, \apj, 553, 922
	
	\bibitem[{{Antiochos} {et~al.}(1999){Antiochos}, {DeVore}, \&
		{Klimchuk}}]{Antiochos1999}
	{Antiochos}, S.~K., {DeVore}, C.~R., \& {Klimchuk}, J.~A. 1999, \apj, 510, 485
	
	\bibitem[{{Awasthi} {et~al.}(2014){Awasthi}, {Jain}, {Gadhiya}, {Aschwanden},
		{Uddin}, {Srivastava}, {Chandra}, {Gopalswamy}, {Nitta}, {Yashiro},
		{Manoharan}, {Choudhary}, {Joshi}, {Dwivedi}, \& {Mahalakshmi}}]{Awasthi2014}
	{Awasthi}, A.~K., {Jain}, R., {Gadhiya}, P.~D., {et~al.} 2014, \mnras, 437,
	2249
	
	\bibitem[{{Burlaga} {et~al.}(1981){Burlaga}, {Sittler}, {Mariani}, \&
		{Schwenn}}]{Burlaga1981}
	{Burlaga}, L., {Sittler}, E., {Mariani}, F., \& {Schwenn}, R. 1981, \jgr, 86,
	6673
	
	\bibitem[{{Burlaga} {et~al.}(2002){Burlaga}, {Plunkett}, \&
		{St.~Cyr}}]{Burlaga2002}
	{Burlaga}, L.~F., {Plunkett}, S.~P., \& {St.~Cyr}, O.~C. 2002, Journal of
	Geophysical Research (Space Physics), 107, 1266
	
	\bibitem[{{Canfield} {et~al.}(1999){Canfield}, {Hudson}, \&
		{McKenzie}}]{Canfield1999}
	{Canfield}, R.~C., {Hudson}, H.~S., \& {McKenzie}, D.~E. 1999, \grl, 26, 627
	
	\bibitem[{{Cheung} {et~al.}(2015){Cheung}, {Boerner}, {Schrijver}, {Testa},
		{Chen}, {Peter}, \& {Malanushenko}}]{Cheung2015}
	{Cheung}, M.~C.~M., {Boerner}, P., {Schrijver}, C.~J., {et~al.} 2015, \apj,
	807, 143
	
	\bibitem[{{Chi} {et~al.}(2016){Chi}, {Shen}, {Wang}, {Xu}, {Ye}, \&
		{Wang}}]{Chi2016}
	{Chi}, Y., {Shen}, C., {Wang}, Y., {et~al.} 2016, \solphys, 291, 2419
	
	\bibitem[{{Chifor} {et~al.}(2007){Chifor}, {Tripathi}, {Mason}, \&
		{Dennis}}]{Chifor2007}
	{Chifor}, C., {Tripathi}, D., {Mason}, H.~E., \& {Dennis}, B.~R. 2007, \aap,
	472, 967
	
	\bibitem[{{De Pontieu} {et~al.}(2014){De Pontieu}, {Title}, {Lemen}, {Kushner},
		{Akin}, {Allard}, {Berger}, {Boerner}, {Cheung}, {Chou}, {Drake}, {Duncan},
		{Freeland}, {Heyman}, {Hoffman}, {Hurlburt}, {Lindgren}, {Mathur}, {Rehse},
		{Sabolish}, {Seguin}, {Schrijver}, {Tarbell}, {W{\"u}lser}, {Wolfson},
		{Yanari}, {Mudge}, {Nguyen-Phuc}, {Timmons}, {van Bezooijen}, {Weingrod},
		{Brookner}, {Butcher}, {Dougherty}, {Eder}, {Knagenhjelm}, {Larsen},
		{Mansir}, {Phan}, {Boyle}, {Cheimets}, {DeLuca}, {Golub}, {Gates}, {Hertz},
		{McKillop}, {Park}, {Perry}, {Podgorski}, {Reeves}, {Saar}, {Testa}, {Tian},
		{Weber}, {Dunn}, {Eccles}, {Jaeggli}, {Kankelborg}, {Mashburn}, {Pust},
		{Springer}, {Carvalho}, {Kleint}, {Marmie}, {Mazmanian}, {Pereira}, {Sawyer},
		{Strong}, {Worden}, {Carlsson}, {Hansteen}, {Leenaarts}, {Wiesmann},
		{Aloise}, {Chu}, {Bush}, {Scherrer}, {Brekke}, {Martinez-Sykora}, {Lites},
		{McIntosh}, {Uitenbroek}, {Okamoto}, {Gummin}, {Auker}, {Jerram}, {Pool}, \&
		{Waltham}}]{DePontieu2014}
	{De Pontieu}, B., {Title}, A.~M., {Lemen}, J.~R., {et~al.} 2014, \solphys, 289,
	2733
	
	\bibitem[{{D{\'e}moulin}(2006)}]{Demoulin2006}
	{D{\'e}moulin}, P. 2006, Advances in Space Research, 37, 1269
	
	\bibitem[{{Forbes}(2000)}]{Forbes2000}
	{Forbes}, T.~G. 2000, \jgr, 105, 23153
	
	\bibitem[{{Forbes} {et~al.}(2006){Forbes}, {Linker}, {Chen}, {Cid}, {K{\'o}ta},
		{Lee}, {Mann}, {Miki{\'c}}, {Potgieter}, {Schmidt}, {Siscoe}, {Vainio},
		{Antiochos}, \& {Riley}}]{Forbes2006}
	{Forbes}, T.~G., {Linker}, J.~A., {Chen}, J., {et~al.} 2006, \ssr, 123, 251
	
	\bibitem[{{Gonzalez} {et~al.}(2007){Gonzalez}, {Echer}, {Clua-Gonzalez}, \&
		{Tsurutani}}]{Gonzalez2007}
	{Gonzalez}, W.~D., {Echer}, E., {Clua-Gonzalez}, A.~L., \& {Tsurutani}, B.~T.
	2007, \grl, 34, L06101
	
	\bibitem[{{Hood} {et~al.}(2009){Hood}, {Archontis}, {Galsgaard}, \&
		{Moreno-Insertis}}]{Hood2009}
	{Hood}, A.~W., {Archontis}, V., {Galsgaard}, K., \& {Moreno-Insertis}, F. 2009,
	\aap, 503, 999
	
	\bibitem[{{Hurford} {et~al.}(2002){Hurford}, {Schmahl}, {Schwartz}, {Conway},
		{Aschwanden}, {Csillaghy}, {Dennis}, {Johns-Krull}, {Krucker}, {Lin},
		{McTiernan}, {Metcalf}, {Sato}, \& {Smith}}]{Hurford2002}
	{Hurford}, G.~J., {Schmahl}, E.~J., {Schwartz}, R.~A., {et~al.} 2002, \solphys,
	210, 61
	
	\bibitem[{{Jing} {et~al.}(2017){Jing}, {Liu}, {Cheung}, {Lee}, {Xu}, {Liu},
		{Zhu}, \& {Wang}}]{Jing2017}
	{Jing}, J., {Liu}, R., {Cheung}, M.~C.~M., {et~al.} 2017, \apjl, 842, L18
	
	\bibitem[{{Jing} {et~al.}(2016){Jing}, {Xu}, {Cao}, {Liu}, {Gary}, \&
		{Wang}}]{Jing2016}
	{Jing}, J., {Xu}, Y., {Cao}, W., {et~al.} 2016, Scientific Reports, 6, 24319
	
	\bibitem[{{Kliem} {et~al.}(2014){Kliem}, {T{\"o}r{\"o}k}, {Titov}, {Lionello},
		{Linker}, {Liu}, {Liu}, \& {Wang}}]{Kliem2014}
	{Kliem}, B., {T{\"o}r{\"o}k}, T., {Titov}, V.~S., {et~al.} 2014, \apj, 792, 107
	
	\bibitem[{{Kosugi} {et~al.}(2007){Kosugi}, {Matsuzaki}, {Sakao}, {Shimizu},
		{Sone}, {Tachikawa}, {Hashimoto}, {Minesugi}, {Ohnishi}, {Yamada}, {Tsuneta},
		{Hara}, {Ichimoto}, {Suematsu}, {Shimojo}, {Watanabe}, {Shimada}, {Davis},
		{Hill}, {Owens}, {Title}, {Culhane}, {Harra}, {Doschek}, \&
		{Golub}}]{Kosugi2007}
	{Kosugi}, T., {Matsuzaki}, K., {Sakao}, T., {et~al.} 2007, \solphys, 243, 3
	
	\bibitem[{{Lemen} {et~al.}(2012){Lemen}, {Title}, {Akin}, {Boerner}, {Chou},
		{Drake}, {Duncan}, {Edwards}, {Friedlaender}, {Heyman}, {Hurlburt}, {Katz},
		{Kushner}, {Levay}, {Lindgren}, {Mathur}, {McFeaters}, {Mitchell}, {Rehse},
		{Schrijver}, {Springer}, {Stern}, {Tarbell}, {Wuelser}, {Wolfson}, {Yanari},
		{Bookbinder}, {Cheimets}, {Caldwell}, {Deluca}, {Gates}, {Golub}, {Park},
		{Podgorski}, {Bush}, {Scherrer}, {Gummin}, {Smith}, {Auker}, {Jerram},
		{Pool}, {Soufli}, {Windt}, {Beardsley}, {Clapp}, {Lang}, \&
		{Waltham}}]{Lemen2012}
	{Lemen}, J.~R., {Title}, A.~M., {Akin}, D.~J., {et~al.} 2012, \solphys, 275, 17
	
	\bibitem[{{Lin} {et~al.}(2002){Lin}, {Dennis}, {Hurford}, {Smith}, {Zehnder},
		{Harvey}, {Curtis}, {Pankow}, {Turin}, {Bester}, {Csillaghy}, {Lewis},
		{Madden}, {van Beek}, {Appleby}, {Raudorf}, {McTiernan}, {Ramaty}, {Schmahl},
		{Schwartz}, {Krucker}, {Abiad}, {Quinn}, {Berg}, {Hashii}, {Sterling},
		{Jackson}, {Pratt}, {Campbell}, {Malone}, {Landis}, {Barrington-Leigh},
		{Slassi-Sennou}, {Cork}, {Clark}, {Amato}, {Orwig}, {Boyle}, {Banks},
		{Shirey}, {Tolbert}, {Zarro}, {Snow}, {Thomsen}, {Henneck}, {McHedlishvili},
		{Ming}, {Fivian}, {Jordan}, {Wanner}, {Crubb}, {Preble}, {Matranga}, {Benz},
		{Hudson}, {Canfield}, {Holman}, {Crannell}, {Kosugi}, {Emslie}, {Vilmer},
		{Brown}, {Johns-Krull}, {Aschwanden}, {Metcalf}, \& {Conway}}]{Lin2002}
	{Lin}, R.~P., {Dennis}, B.~R., {Hurford}, G.~J., {et~al.} 2002, \solphys, 210,
	3
	
	\bibitem[{{Linton} {et~al.}(2001){Linton}, {Dahlburg}, \&
		{Antiochos}}]{Linton2001}
	{Linton}, M.~G., {Dahlburg}, R.~B., \& {Antiochos}, S.~K. 2001, \apj, 553, 905
	
	\bibitem[{{Liu} {et~al.}(2016{\natexlab{a}}){Liu}, {Chen}, {Wang}, \&
		{Liu}}]{Liu2016SR}
	{Liu}, R., {Chen}, J., {Wang}, Y., \& {Liu}, K. 2016{\natexlab{a}}, Scientific
	Reports, 6, 34021
	
	\bibitem[{{Liu} {et~al.}(2012){Liu}, {Kliem}, {T{\"o}r{\"o}k}, {Liu}, {Titov},
		{Lionello}, {Linker}, \& {Wang}}]{Liu2012}
	{Liu}, R., {Kliem}, B., {T{\"o}r{\"o}k}, T., {et~al.} 2012, \apj, 756, 59
	
	\bibitem[{{Liu} {et~al.}(2010){Liu}, {Liu}, {Wang}, {Deng}, \&
		{Wang}}]{Liu2010}
	{Liu}, R., {Liu}, C., {Wang}, S., {Deng}, N., \& {Wang}, H. 2010, \apjl, 725,
	L84
	
	\bibitem[{{Liu} {et~al.}(2014){Liu}, {Titov}, {Gou}, {Wang}, {Liu}, \&
		{Wang}}]{Liu2014}
	{Liu}, R., {Titov}, V.~S., {Gou}, T., {et~al.} 2014, \apj, 790, 8
	
	\bibitem[{{Liu} {et~al.}(2016{\natexlab{b}}){Liu}, {Kliem}, {Titov}, {Chen},
		{Wang}, {Wang}, {Liu}, {Xu}, \& {Wiegelmann}}]{Liu2016}
	{Liu}, R., {Kliem}, B., {Titov}, V.~S., {et~al.} 2016{\natexlab{b}}, \apj, 818,
	148
	
	\bibitem[{{Liu} {et~al.}(2013){Liu}, {Luhmann}, {Lugaz}, {M{\"o}stl}, {Davies},
		{Bale}, \& {Lin}}]{Liu2013}
	{Liu}, Y.~D., {Luhmann}, J.~G., {Lugaz}, N., {et~al.} 2013, \apj, 769, 45
	
	\bibitem[{{Loureiro} \& {Uzdensky}(2016)}]{Loureiro&Udensky2016}
	{Loureiro}, N.~F., \& {Uzdensky}, D.~A. 2016, Plasma Physics and Controlled
	Fusion, 58, 014021
	
	\bibitem[{{Lynch} {et~al.}(2008){Lynch}, {Antiochos}, {DeVore}, {Luhmann}, \&
		{Zurbuchen}}]{Lynch2008}
	{Lynch}, B.~J., {Antiochos}, S.~K., {DeVore}, C.~R., {Luhmann}, J.~G., \&
	{Zurbuchen}, T.~H. 2008, \apj, 683, 1192
	
	\bibitem[{{Marscher} {et~al.}(2008){Marscher}, {Jorstad}, {D'Arcangelo},
		{Smith}, {Williams}, {Larionov}, {Oh}, {Olmstead}, {Aller}, {Aller},
		{McHardy}, {L{\"a}hteenm{\"a}ki}, {Tornikoski}, {Valtaoja}, {Hagen-Thorn},
		{Kopatskaya}, {Gear}, {Tosti}, {Kurtanidze}, {Nikolashvili}, {Sigua},
		{Miller}, \& {Ryle}}]{Marscher2008}
	{Marscher}, A.~P., {Jorstad}, S.~G., {D'Arcangelo}, F.~D., {et~al.} 2008, \nat,
	452, 966
	
	\bibitem[{{Moore} \& {Sterling}(2006)}]{Moore2006}
	{Moore}, R.~L., \& {Sterling}, A.~C. 2006, Washington DC American Geophysical
	Union Geophysical Monograph Series, 165, 43
	
	\bibitem[{{Moore} {et~al.}(2001){Moore}, {Sterling}, {Hudson}, \&
		{Lemen}}]{Moore2001}
	{Moore}, R.~L., {Sterling}, A.~C., {Hudson}, H.~S., \& {Lemen}, J.~R. 2001,
	\apj, 552, 833
	
	\bibitem[{{Parker}(1983)}]{Parker1983a}
	{Parker}, E.~N. 1983, \apj, 264, 635
	
	\bibitem[{{Parker}(1988)}]{Parker1988}
	---. 1988, \apj, 330, 474
	
	\bibitem[{{Priest} \& {Forbes}(2000)}]{Priest&Forbes2000}
	{Priest}, E., \& {Forbes}, T. 2000, {Magnetic Reconnection: MHD theory and
		applications} (Cambridge University Press), 612
	
	\bibitem[{{Rust}(2003)}]{Rust2003}
	{Rust}, D.~M. 2003, Advances in Space Research, 32, 1895
	
	\bibitem[{{Rust} \& {Kumar}(1996)}]{Rust&Kumar1996}
	{Rust}, D.~M., \& {Kumar}, A. 1996, \apjl, 464, L199
	
	\bibitem[{{Scherrer} {et~al.}(2012){Scherrer}, {Schou}, {Bush}, {Kosovichev},
		{Bogart}, {Hoeksema}, {Liu}, {Duvall}, {Zhao}, {Title}, {Schrijver},
		{Tarbell}, \& {Tomczyk}}]{Scherrer2012}
	{Scherrer}, P.~H., {Schou}, J., {Bush}, R.~I., {et~al.} 2012, \solphys, 275,
	207
	
	\bibitem[{{Shen} {et~al.}(2017){Shen}, {Chi}, {Wang}, {Xu}, \&
		{Wang}}]{Shen2017}
	{Shen}, C., {Chi}, Y., {Wang}, Y., {Xu}, M., \& {Wang}, S. 2017, Journal of
	Geophysical Research (Space Physics), 122, 5931
	
	\bibitem[{{Su} {et~al.}(2011){Su}, {Surges}, {van Ballegooijen}, {DeLuca}, \&
		{Golub}}]{Su2011}
	{Su}, Y., {Surges}, V., {van Ballegooijen}, A., {DeLuca}, E., \& {Golub}, L.
	2011, \apj, 734, 53
	
	\bibitem[{{Takasao} {et~al.}(2012){Takasao}, {Asai}, {Isobe}, \&
		{Shibata}}]{Takasao2012}
	{Takasao}, S., {Asai}, A., {Isobe}, H., \& {Shibata}, K. 2012, \apjl, 745, L6
	
	\bibitem[{{Taylor}(1986)}]{Taylor1986}
	{Taylor}, J.~B. 1986, Reviews of Modern Physics, 58, 741
	
	\bibitem[{{Titov} {et~al.}(2002){Titov}, {Hornig}, \&
		{D{\'e}moulin}}]{Titov2002}
	{Titov}, V.~S., {Hornig}, G., \& {D{\'e}moulin}, P. 2002, Journal of
	Geophysical Research (Space Physics), 107, 1164
	
	\bibitem[{{Vourlidas} {et~al.}(2013){Vourlidas}, {Lynch}, {Howard}, \&
		{Li}}]{Vourlidas2013}
	{Vourlidas}, A., {Lynch}, B.~J., {Howard}, R.~A., \& {Li}, Y. 2013, \solphys,
	284, 179
	
	\bibitem[{{Wang} {et~al.}(2017{\natexlab{a}}){Wang}, {Liu}, {Ahn}, {Xu},
		{Jing}, {Deng}, {Huang}, {Liu}, {Kusano}, {Fleishman}, {Gary}, \&
		{Cao}}]{Wang2017}
	{Wang}, H., {Liu}, C., {Ahn}, K., {et~al.} 2017{\natexlab{a}}, Nature
	Astronomy, 1, 0085
	
	\bibitem[{{Wang} {et~al.}(2017{\natexlab{b}}){Wang}, {Liu}, {Wang}, {Hu},
		{Shen}, {Jiang}, \& {Zhu}}]{Wang2017nc}
	{Wang}, W., {Liu}, R., {Wang}, Y., {et~al.} 2017{\natexlab{b}}, Nature
	Communications, 8, 1330
	
	\bibitem[{{Wang} {et~al.}(2003){Wang}, {Ye}, \& {Wang}}]{Wang2003}
	{Wang}, Y.~M., {Ye}, P.~Z., \& {Wang}, S. 2003, Journal of Geophysical Research
	(Space Physics), 108, 1370
	
	\bibitem[{{Webb} \& {Howard}(2012)}]{Webb&Howard2012}
	{Webb}, D.~F., \& {Howard}, T.~A. 2012, Living Reviews in Solar Physics, 9, 3
	
	\bibitem[{{Wiegelmann} {et~al.}(2006){Wiegelmann}, {Inhester}, {Kliem},
		{Valori}, \& {Neukirch}}]{Wiegelmann2006}
	{Wiegelmann}, T., {Inhester}, B., {Kliem}, B., {Valori}, G., \& {Neukirch}, T.
	2006, \aap, 453, 737
	
	\bibitem[{{Zhang} {et~al.}(2012){Zhang}, {Cheng}, \& {Ding}}]{Zhang2012}
	{Zhang}, J., {Cheng}, X., \& {Ding}, M.-D. 2012, Nature Communications, 3, 747
	
	\bibitem[{{Zhang} {et~al.}(2014){Zhang}, {Liu}, {Wang}, {Shen}, {Liu}, {Liu},
		\& {Wang}}]{Zhang2014}
	{Zhang}, Q., {Liu}, R., {Wang}, Y., {et~al.} 2014, \apj, 789, 133
	
	\bibitem[{{Zhu} {et~al.}(2015){Zhu}, {Liu}, {Alexander}, {Sun}, \&
		{McAteer}}]{Zhu2015}
	{Zhu}, C., {Liu}, R., {Alexander}, D., {Sun}, X., \& {McAteer}, R.~T.~J. 2015,
	\apj, 813, 60
	
	\bibitem[{{Zurbuchen} \& {Richardson}(2006)}]{Zurbuchen&Richardson2006}
	{Zurbuchen}, T.~H., \& {Richardson}, I.~G. 2006, \ssr, 123, 31
	
\end{thebibliography}

\end{document}